# Tunneling and Percolation in Metal-Insulator Composite Materials

D. Toker, D. Azulay, N. Shimoni, I. Balberg, and O. Millo*

The Racah Institute of Physics, The Hebrew University, Jerusalem 91904, Israel

Abstract

In many composites the electrical transport takes place only by tunneling between isolated particles. For a long time it was quite a puzzle how, in spite of the incompatibility of tunneling and percolation networks, these composites conform well to percolation theory. We found, by conductance atomic force microscopy measurements on granular metals, that it is the apparent cut-off of the tunneling to non-nearest neighbors that brings about this behavior. In particular, the percolation cluster is shown to consist of the nearest-neighbors sub-network of the full tunneling network.

*Corresponding author: Oded Millo, milode@vms.huji.ac.il

The electrical conduction in composite systems comprising an intricate network of conducting and insulating phases is determined by two mechanisms, percolation in a continuous conducting network[1] and/or tunneling[2,3] between isolated conducting particles (grains, crystallites, etc.). In the pioneering works on such systems, these two mechanisms have been considered separately.[1,2,3,4,5,6] In particular, for a high enough content of the metallic phase in granular metals, the continuous network is formed by the coalescence of the metallic grains. This continuity yields, as expected from "classical" percolation theory,[7] the typical power-law dependence of the electrical conductivity, $\sigma$, on the vol. % of the conducting phase, x:

$$\sigma \propto (x-x_c)^t, \qquad (1)$$

where t is a corresponding critical exponent. The value of $x_c$ that was derived by fitting[1,5] Eq. (1) to the experimental data on granular metals is in excellent agreement with that extracted from structural and other physical data,[4,5] which show the formation of a continuous metallic network. Moreover, the conductivity exponent was found to be very close to the predicted[7] universal value of t = 2.0, confirming that the value derived experimentally for $x_c$ is exactly the (global geometrical connectivity) threshold that is considered in percolation theory.[7,8] On the other hand, in contrast with the expectation from percolation theory,[7] a considerable finite conductivity was observed in the regime of $x < x_c$ (the so-called dielectric regime[4,5]), where a continuous metallic network is absent. This conductivity was attributed to interparticle tunneling.[2,4,9,10] However, while the theories proposed to account for the electrical conductance in that regime have considered the tunneling within a single pair of adjacent particles, the global electrical connectedness of the tunneling conduction network has not been discussed. This approach was also carried over to other composites,[3,6] where the particles do not coalesce or fuse[11] for any x.

While a priori justified, the latter approach has been ignored by numerous authors, and transport in systems lacking a continuous geometrical network of the conducting phase, such as carbon black-polymer[12,13,14] and other[15,16] composites, were analyzed in terms of percolation theory. Surprisingly, excellent agreement was found[12,13,14,15,16] between corresponding experimental results and the predictions of percolation theory. In particular, agreement was found, by applying Eq. (1), with universal[12] or non-universal[17,18] values of t. However, the meaning and the values of $x_c$ that were found from these fits were not discussed, and the applicability of the above phenomenological percolation approach has not yet been justified.

When attempting to find such a justification and the meaning of $x_c$, one immediately notes that a system of separated conducting particles with tunneling conduction is a priori incompatible[19] with the percolation picture. Geometrically, no two particles are in contact (and then no well-defined percolation cluster and percolation threshold are apparent), and electrically all particles are in contact with each other by tunneling (and then no percolation transition in the electrical properties is expected). In the composites considered, we have then a homogeneous network of tunneling conductors to which we refer here as all-connected tunneling network (ACTN). The question that arises then is how is it that, in spite of this global tunneling connectivity of the network, the electrical conduction in corresponding composites is governed by the very well-defined percolation-like behavior.[14,19] The purpose of the present letter is to resolve this issue, thus providing an a posteriori justification for applying[12,13,14,15,16,17,18] the percolation approach to the analysis of non-continuous conducting networks. In particular, we will show that $x_c$ has a well defined meaning in such systems, and argue that the ACTN can be reduced to a well-defined percolation network, such as in porous media[20] where the geometrical and electrical networks coincide. To that end, one would like to obtain local current maps of these systems, or a geometrical-statistical property that can account for them, such as the fractal dimension of the current-carrying percolation cluster.

Following the success of the application of conductance atomic force microscopy (C-AFM) to granular metals[21] and our ability[22] to derive experimentally (with high accuracy) that fractal dimension of the percolation cluster in carbon black-polymer composites (CBPC), we have applied this method to the present study of granular metals.

The basic idea of this work on Ni-SiO$_2$ composites is then to discriminate between various current maps by their fractal dimension. As we show below, this approach is found useful for providing an answer to the above questions, for granular metals in the $x < x_c$ regime in particular, and for two-phase composites where the conducting particles are geometrically separated in general.

The Ni-SiO$_2$ films, 2.5-5.5 μm thick and 5 mm wide, were deposited as we described previously[23] on Pyrex substrates by co-sputtering of Ni and SiO$_2$. The co-sputtering method[4,5] enables a continuous variation of the metal content (vol.% Ni) along the deposited film, where one end is Ni rich and the other end is insulator rich. The typical size of the Ni particles increases continuously from 4 nm at x = 35 vol.% to 10 nm at x = 50 vol.%, to a continuous network at x = 90 vol.%.[4,5,23,24] For the electrical measurements Cu-Au electrodes were predeposited on the substrate, with a separation of 2.54 mm. The resistance along the substrate sections was determined by a standard four-probe technique. For the C-AFM measurements some of the samples were cut out, leaving on each sample an electrode that served as a counter-electrode to the scanning conducting AFM tip. For these measurements we have used samples with Ni contents of 24, 28, 35, 47, 60, 72, 80, 87 and 92 vol.%. The details of our C-AFM apparatus and the procedure of our measurements were presented recently[22] and will not be repeated here. We point out, however, that the spatial resolution of the current images was limited only by the pixel size (typically 10x10 nm$^2$) and that, since the tip was scanning more than 1 mm from the counter electrode, the measured current images reflect two-dimensional cross-sections of the three-dimensional macroscopic conduction network. We refer to areas of the image where current from the tip to the counter-electrode is detected as "conducting islands", and the lowest current that we consider in the analysis of an image is denoted as the "cut-off current", I$_{co}$.[22] Obviously, the total area covered by the conducting islands, A$_t$, depends on the chosen I$_{co}$. The larger the considered I$_{co}$ the smaller the A$_t$, since only lower resistance paths of the conduction network are included.

We point out that no correlation was found between the AFM topographic images and the C-AFM images, implying that the current maps reflect mainly the electrical properties of the bulk of the sample.[22]

To characterize our samples we show in Fig. 1 the dependence of the conductivity on the Ni vol.%. These results are much the same as those obtained on many granular metals.[1,4,5,25] For x values beyond a certain critical value, $x_c$, the electrical conductivity follows the expected universal percolation behavior with t ≈ 2, as shown by the corresponding fit to Eq. (1). The value of $x_c$ = 52 vol.% found here is in excellent agreement with previous electrical and structural measurements performed on this system.[5,23,24,25] The non-negligible "excess" conductivity in the $x < x_c$ regime is attributed, as mentioned above, to inter-grain tunneling[2,4,5,26] and it is that regime which is of interest here.

Turning to our C-AFM measurements, we show in the insets of Fig. 2(a) typical current images for a sample of 35 vol.% Ni for two values of $I_{co}$, 0.05 and 0.5 nA, in the upper and lower insets, respectively. It is clearly seen that $A_t$ increases with the decrease of $I_{co}$. For finding the fractal dimension of the three-dimensional percolation cluster $D_3$, we recall[22] that $D_3$ = D+1, where D is the fractal dimension of $A_t$. The determination of D is carried out by measuring $A_t$ as a function of the "window length" L. In this procedure we considered L values that are much larger than the particle size but much smaller than the correlation length expected for a percolation network in the vicinity of $x_c$ (tens of microns). For the latter regime one expects the scaling behavior,[7] $A_t \propto L^D$, with D=1.5 ($D_3$ = 2.5). On the other hand, as one departs from the percolation threshold, $D_3$ should approach the homogeneous, Euclidian dimension of space, i.e. 3. The two-dimensional images should yield accordingly a value of D = 1.5 in the former (x just above $x_c$) regime and a value of D = 2 in the latter (x >> $x_c$) regime. Indeed we have previously found this behavior for the CBPC system.[22] The corresponding results of our present study on Ni-SiO$_2$ samples are shown in Fig. 2. The behavior in Fig. 2(a) for $x < x_c$ reveals conductance associated with the network of tunneling transport in the sample.

The new observation, in comparison with the findings on CBPC, is that the fractal dimension, D, decreases with the increase of the $I_{co}$ value considered in the analysis. This effect is beyond the experimental error or the statistical error (of the log-log-fit) that is indicated in the figure. The important consequence is that the type of the current network that participates for each $I_{co}$ value is *different* and that for the high $I_{co}$ values, only inter-grain tunneling resistors that are lower than a given value are included in the observed conducting network. The high $I_{co}$ images thus correspond statistically to the contribution of the lowest resistors of the network, i.e. to currents that transverse the sample via a route that consists mainly of particles that are separated by the typical tunneling distance (here, "nearest neighbors"). In contrast, the low $I_{co}$ values include current paths that are added to the above inter-adjacent grain paths. These additional paths correspond to larger (tunneling) resistors that are either removed "nearest neighbors" or originate from higher order neighbors. Hence, the network defined by the highest $I_{co}$ values is similar to the one encountered in fused-particles[1,5] (or porous media-like) composites,[8,27] namely, a bona fide percolation-like network. In contrast, the very low $I_{co}$ values correspond to an ACTN that is homogeneous and thus the fractal dimension of the current paths network approaches the regular dimension of the system (D = 2, $D_3$ = 3).

The above picture is further supported by the results that are shown in Fig. 2(b) for x just above $x_c$. Here, we expect a priori the fractal dimension of the connected continuous metallic network, i.e. $D_3$ = 2.5. We note however, that if we also include now the tunneling contribution in this regime, as was previously suggested by other authors,[26] a homogenous tunneling component will be added to the percolation system of the continuous metallic network. Hence the value of $D_3$ should increase towards $D_3$ = 3 with the decrease of $I_{co}$, as observed. The whole framework of our explanations is validated by the results shown in Fig. 2(c). Here we see that far above the percolation threshold (x = 92 vol%) where a genuine homogeneous-continuous network exists, the observed dimension is essentially 3, independent of the value of $I_{co}$. For comparison, let us mention that for the CBPC the value of $D_3$ was found to change from 2.5, in the vicinity of $x_c$, to 3, for $x > x_c$, but *no dependence* on $I_{co}$ has been observed.[22]

Another, independent, observation that supports the above interpretation is derived from the bias dependence of the fractal dimension, which is found to increase monotonically with the tip's bias. For example, for $x = 35$ vol.% and $I_{co} = 0.05$ nA, the value of $D_3$ was found to increase from 1.54 at 2V to 1.73 at 3.1 V. Indeed, further conducting channels, involving tunneling between particles with larger interparticle distances, are expected to open-up with the increase of bias.[26] These channels add to the conduction paths in the ACTN, resulting in a more homogenous network, and consequently a higher value of D (and of $D_3$).

Our interpretation of the above results explains well now why the percolation predicted[7] fractal dimension has been observed[22] for the CBPC *regardless* of the value of $I_{co}$. In contrast with the small effective radius of the conducting particles (and then the relatively small distances to particles beyond the nearest neighbor[2,5]) in the present granular metal systems, R(GM), the effective radius of the carbon particles, R(CB) (~100 nm), is very large in comparison with the typical tunneling distance d (a few nm).[3,6,17] Hence, the interparticle surface-to-surface distance of the "next nearest neighbors" is so much longer in comparison with the corresponding distance of the first nearest neighbors (which is of the order[13,17] of d) that the conduction between a given particle and a non nearest-neighbor is negligible (even in comparison with the conduction in the "insulating" polymer matrix). This picture explains then our $D_3 = 2.5$-*only* finding as well as the "pure" percolation-like behavior (as manifested by agreement with Eq. (1)) that was generally observed for the CBPC.[12,13] The fact that a "geometrical" percolation-like correlation length was found[14] in CBPC is also well explained now. This is since the correlation length was determined by an electrical, rather than a structural, measurement, and thus, as explained above, the corresponding electrical system is a percolation cluster-like system. We see then that for all practical purposes the CBPC conduct only via nearest neighbor tunneling. The CBPC constitute then, as do the porous materials,[20,27] bona fide percolation systems. The above interpretation is illustrated in Fig. 3 where we show two small ACTN systems in which the interparticle tunneling connections are indicated by connecting segments. The thick segments correspond to "nearest neighbor" tunneling connections and the thin segments to "higher neighbor" tunneling connections. In Fig. 3(a) we illustrate the network in the granular metals and in Fig. 3(b) the network in the CBPC. In the former case the thick segments represent the effective network only for high $I_{co}$ values while in the latter case they are practically the only ones that are effective, for any value of $I_{co}$. The above interpretation also explains the meaning of $x_c$ in all composite systems where the transport is carried by interparticle tunneling. The experimentally determined value of $x_c$ is the percolation threshold of an electrical network in which *only* the tunneling between adjacent particles contributes to the conduction.

Following all the above we can conclude that our C-AFM measurements discriminate between various conducting networks. In particular, when the contribution to the conduction is only due to nearest neighbor tunneling the system behaves as a genuine percolation system. This explains the "pure" percolation behavior that is observed in many composites where the conducting particles are geometrically separated from each other and where they are much larger than the tunneling range.

*Acknowledgement*: This work was supported in part by the Israel Science Foundation.

Fig. 1 The dependence of the conductivity of our Ni-SiO2 composites on the metallic content (in vol.% of Ni). The curve represents a fit of the data above the geometrical percolation threshold, $x_c$, to Eq. (1).

Fig. 2. The dependence of the fractional area covered by the "conducting islands" on the "window's length", for a few values of the current cut-off: (a) for a sample of x = 35 Ni vol.%, and an applied tip bias of V = 2V, (b) x = 60 vol.%, V = 1.6 V, (c) x = 92 vol.%, V = 1.5 V. The insets show 5x5 $\mu m^2$ current images measured on the x=35 vol.% sample, for two values of the currant cut-off, 0.05 nA (upper inset) and 0.5nA (lower inset). The maximum current in both images is 5 nA.

Fig. 3. An illustration of small parts of an ACTN. The gray circles represent the conducting particles and their white shells represent the effective tunneling range, d. The thick segments represent the tunneling connections between "nearest neighbors" while the thin segments represent tunneling connections between higher order neighbors. The relation between the size of the particles and the tunneling range are illustrated for granular metals in (a) and for carbon black polymer composites in (b).

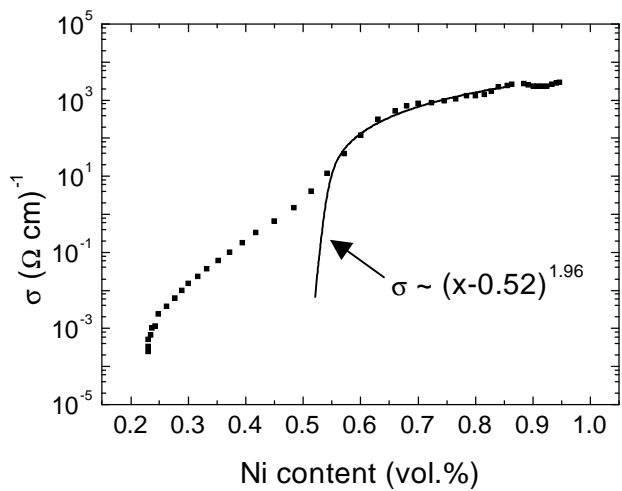

Fig 1, Toker et al.

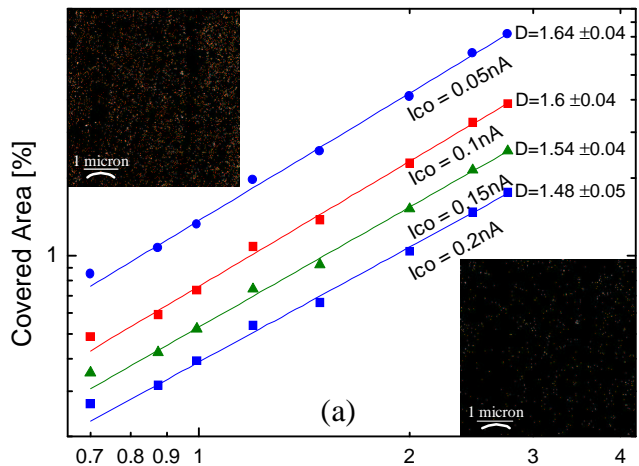

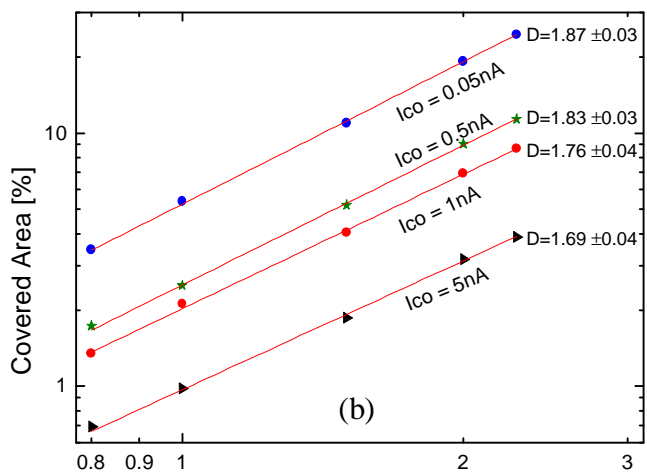

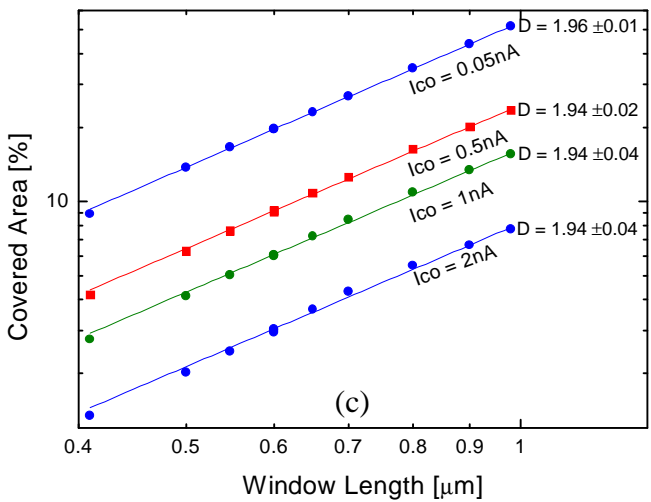

Fig 2, Toker et al.

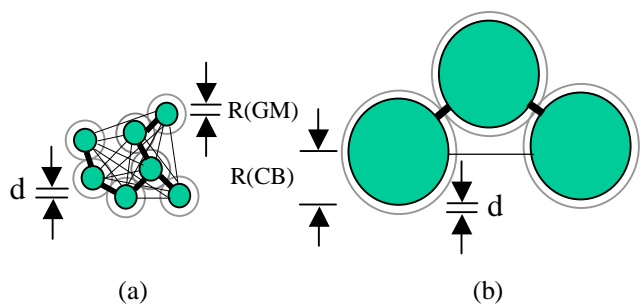

Fig 3, Toker et al.